\documentclass[conference]{IEEEtran}
\IEEEoverridecommandlockouts

\usepackage[style=ieee,minnames=1,maxnames=100,maxcitenames=2]{biblatex} 
\usepackage{amsmath,amssymb,amsfonts}
\usepackage[inline]{enumitem}
\usepackage{graphicx}
\usepackage{textcomp}
\usepackage{xcolor}
\usepackage{subcaption}
\usepackage{multirow}
\usepackage{tcolorbox}
\usepackage{listings}
\usepackage{balance}

\definecolor{codegreen}{rgb}{0,0.6,0}
\definecolor{codegray}{rgb}{0.5,0.5,0.5}
\definecolor{codepurple}{rgb}{0.58,0,0.82}
\definecolor{backcolour}{rgb}{0.95,0.95,0.92}
\lstdefinestyle{mystyle}{
    commentstyle=\color{codegreen},
    keywordstyle=\color{magenta},
    stringstyle=\color{codepurple},
    basicstyle=\ttfamily\scriptsize,
    breakatwhitespace=false,         
    breaklines=true,                 
    keepspaces=true,   
    numbers=left,
    numbersep=1mm, 
    xleftmargin=2mm,
    showspaces=false,                
    showstringspaces=false,
    showtabs=false,                  
    tabsize=2,
    aboveskip=-2.5mm,
    belowskip=-3mm
}

\begin{document}

\renewcommand{\arraystretch}{1.25}

\title{
Predicting Line-Level Defects by Capturing Code Contexts with Hierarchical Transformers
}

\author{
\IEEEauthorblockN{Parvez Mahbub}
\IEEEauthorblockA{
\textit{Faculty of Computer Science} \\
\textit{Dalhousie University, Canada}\\
parvezmrobin@dal.ca}
\and
\IEEEauthorblockN{Mohammad Masudur Rahman}
\IEEEauthorblockA{
\textit{Faculty of Computer Science} \\
\textit{Dalhousie University, Canada}\\
masud.rahman@dal.ca}
}

\maketitle

\begin{abstract}
Software defects consume 40\% of the total budget in software development and cost the global economy billions of dollars every year. 
Unfortunately, despite the use of many software quality assurance (SQA) practices in software development (e.g., code review, continuous integration), defects may still exist in the official release of a software product. 
Therefore, prioritizing SQA efforts for the vulnerable areas of the codebase is essential to ensure the high quality of a software release. 
Predicting software defects at the line level could help prioritize the SQA effort but is a highly challenging task given that only $\approx$~3\% lines of a codebase could be defective. 
Existing works on line-level defect prediction often fall short and cannot fully leverage the line-level defect information. 
In this paper, we propose – Bugsplorer – a novel deep-learning technique for line-level defect prediction. 
It leverages a hierarchical structure of transformer models to represent two types of code elements: code tokens and code lines. 
Unlike the existing techniques that are optimized for file-level defect prediction, Bugsplorer is optimized for a line-level defect prediction objective. 
Our evaluation with five performance metrics shows that Bugsplorer has a promising capability of predicting defective lines with 26-72\% better accuracy than that of the state-of-the-art technique. It can rank the first 20\% defective lines within the top 1-3\% suspicious lines. Thus, Bugsplorer has the potential to significantly reduce SQA costs by ranking defective lines higher.
\end{abstract}


\section{Introduction}
\looseness=-1
A software defect is an erroneous step, process, or data definition in a computer program~\cite{ieee-standard}.
Defect (a.k.a. bug) resolution is one of the major challenges of software development and maintenance. 
According to several studies, it consumes up to 40\% of the total budget~\cite{Glass2001} and costs the global economy billions of dollars each year~\cite{Britton2013, zou2018practitioners}.
Software Quality Assurance (SQA) practices play a critical role in preventing these defects.
However, despite using many SQA practices in the development phase (e.g., code review, continuous integration), defects may still exist in the official release of a software product~\cite{thongtanunam2015investigating, thongtanunam2016revisiting}.
Besides, according to a recent study~\cite{pornprasit2022deeplinedp}, only $\approx$~3\% lines of code from the whole release could lead to most of the defects.
Therefore, prioritizing SQA efforts for the vulnerable areas of the code base is essential to ensure the high quality of a software release.

\looseness=-1
Defect prediction has been a popular research topic for the last few decades. 
It predicts potential defects in software code, which could be useful to improve the software quality, before releasing the product to end users.
It can also help prioritize the SQA efforts.
Defects can be predicted at various abstraction levels of code such as module~\cite{gong2021revisiting, yu2019empirical}, file~\cite{jiarpakdee2021practitioners, chen2020software}, method~\cite{shippey2019automatically}, and line~\cite{wattanakriengkrai2020predicting, pornprasit2021jitline, pornprasit2022deeplinedp}.
Among them, line-level defect prediction provides the most fine-grained location of a software defect, which can reduce the effort to address the defect.

The majority of the contemporary approaches for line-level defect prediction first train their machine learning models to predict the defective source files~\cite{wattanakriengkrai2020predicting, pornprasit2022deeplinedp} or commits~\cite{pornprasit2021jitline}.
Then, if a file or commit is predicted as defective, they identify the tokens in the file that help explain the defects using various techniques (e.g., attention mechanism~\cite{vaswani2017attention}).
Finally, they mark such lines of code from the source file as defective that contain many defect-explaining tokens.
However, such an approach poses two major challenges as follows.

\begin{figure}
    \centering
    \includegraphics[width=\linewidth]{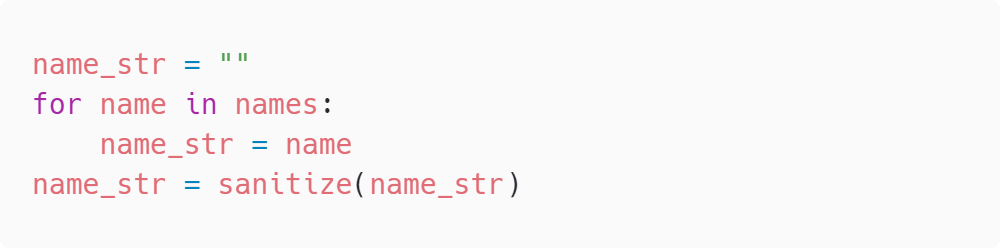}
    \caption{An example of defective code}
    \label{fig: buggy-code}
    \vspace{-1em}
\end{figure}

\paragraph{Existing models might not optimally represent code elements}
The surrounding tokens from both sides could influence the meaning and intent of a code token.
For example, Fig.~\ref{fig: buggy-code} shows a piece of defective code, where a code token -- \texttt{name\_str} -- contains an erroneous value after the program execution.
That is, inside the for loop, the variable \texttt{name} should be concatenated (i.e., \texttt{+=} operator) to \texttt{name\_str} instead of being assigned (i.e., \texttt{=} operator).
Therefore, the code token \texttt{name\_str} was led to be buggy by another code token, ``\texttt{=}'', which appeared later.
The intent of the token \texttt{name\_str} is also influenced by the earlier tokens, such as the token \texttt{for}, by repeating the assignment operation multiple times.
Such a phenomenon indicates that we need information on the surrounding tokens from both sides to represent a token optimally.
However, the technique used in existing study~\cite{pornprasit2022deeplinedp} (e.g., RNN) can only focus on a single direction (a.k.a.\ unidirectional), which could be either earlier tokens or later tokens.
Then, they concatenate two unidirectional representations of a token's context to generate a bidirectional representation.
However, \textcite{reimers2019sentence} suggest that simple concatenation of two vectors might not produce an optimal representation for an input (e.g., a token or line).

\begin{figure*}
    \includegraphics[width=\linewidth]{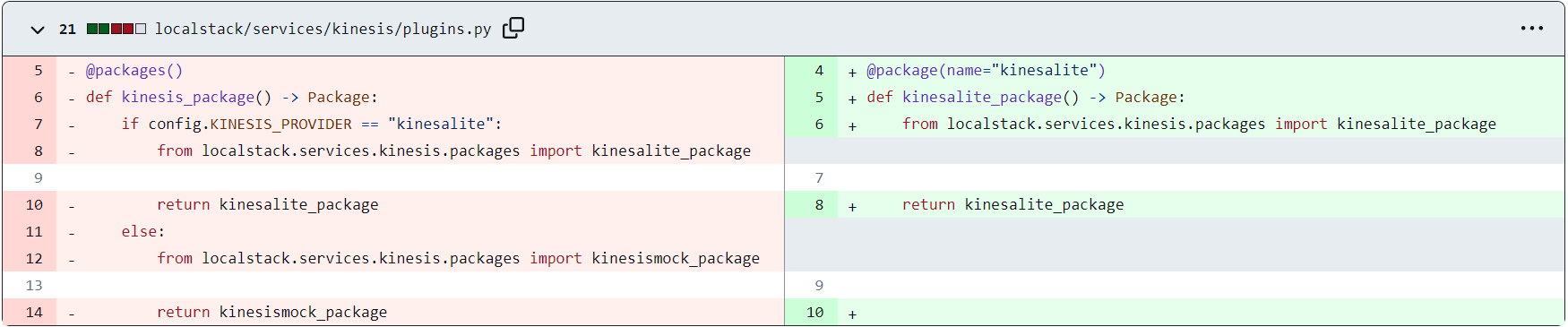}
    \caption{Motivating example for Bugsplorer}
    \label{fig: bugsplorer-motivate}
    \vspace{-1em}
\end{figure*}

\paragraph{Existing models might fail to capture the local context of a defect}
During the training of the models from existing works~\cite{wattanakriengkrai2020predicting, pornprasit2022deeplinedp, pornprasit2021jitline}, the attention values~\cite{bahdanau2014neural} for the tokens are optimized for file-level defect prediction.
In other words, these values are optimized to predict whether the whole file is defective or not.
However, source code documents are often quite large, containing thousands of tokens, which could make them noisy.
Therefore, the attention values from existing models might fail to properly capture the local context of a software defect since, in a codebase, only $\approx$~3\% lines could be defective~\cite{pornprasit2022deeplinedp}.
Thus, relying on these attention values might not be sufficient to detect line-level defects accurately.

In this paper, we propose -- \textit{Bugsplorer} -- a novel deep-learning technique for line-level defect prediction. 
It leverages two transformer models in a hierarchical structure to estimate the attention values for two types of code elements: code tokens and code lines.
Our solution can address the above challenges, which makes our work \emph{novel}. 
First, unlike sequential models (e.g., RNN), Bugsplorer can learn the representation of a code element (e.g., token or line) by capturing its context from both earlier and later tokens simultaneously.
Second, unlike existing techniques~\cite{wattanakriengkrai2020predicting, pornprasit2022deeplinedp, pornprasit2021jitline}, Bugsplorer is directly trained for line-level defect prediction and thus can better capture the local context of a software defect.
Thus, our approach is better suited to predict the line-level defects.

We train and evaluate Bugsplorer with two benchmark datasets -- Defectors~\cite{mahbub2023defectors} and LineDP~\cite{wattanakriengkrai2020predicting}.
The first dataset consists of $\approx$~230K Python source code documents from 24 GitHub repositories.
The second dataset consists of 32 releases spanning 9 Java software systems.
We find that Bugsplorer can predict defective code lines with 26-68\% higher accuracy than that of the state-of-the-art technique~\cite{pornprasit2022deeplinedp}.
It can also reduce the effort in finding defective lines by 72-81\%.
Through an ablation study, we further show that (a) the optimization of deep learning models for line-level defect prediction and (b) the use of bidirectional representations for code elements (e.g., tokens and lines) can significantly influence the performance of our technique.

\vspace{1em}
We thus make the following contribution in this study.
\begin{enumerate}[label=(\alph*)]
\item A novel technique -- Bugsplorer, for line-level defect prediction leveraging hierarchically structured transformers.
\item A comprehensive evaluation and validation of Bugsplorer in terms of both classification and cost-effectiveness metrics using two benchmark datasets: Defectors (24 Python systems)~\cite{mahbub2023defectors} and LineDP (9 Java systems)~\cite{wattanakriengkrai2020predicting}.
\item A replication package (as supplementary material) that includes our working prototype and other configuration details for the replication or third-party reuse.
\end{enumerate}

\section{Motivating Example}

To demonstrate the capability of our technique -- Bugsplorer, let us consider the example in Fig.~\ref{fig: bugsplorer-motivate}.
The code snippet is taken from the \texttt{ray-project/ray} repository at GitHub\footnote{https://bit.ly/3N1NSOf}.
The buggy code attempts to return the driver for the \emph{Amazon Kinesis} service based on configuration.
In particular, it returns the \texttt{kinesalite} driver if it is explicitly specified in the configuration and the \texttt{kinesismock} package otherwise.
Here, the bug is that the \texttt{kinesismock} package should only be used during testing, but this function returns the package even outside the testing environment when no configuration is available.
Therefore, the defect can be found in two places.
The first one is on line 7, where the configuration is checked with an \texttt{if} condition.
The second one is on line 10 and line 14, where an incorrect value is returned from the function.
Bugsplorer can rank all of these defective lines within the first percentile (85th, 83rd, and 79th positions, respectively).
On the other hand, the state-of-the-art technique -- DeepLineDP~\cite{pornprasit2022deeplinedp} -- ranks these three lines beyond the 50th and 80th percentiles, which are much lower in the ranked list.

DeepLineDP is trained with a file-level defect prediction objective, and thus its focus is intuitively scattered over the whole file.
As a result, it might fail to precisely capture the local context of the defect.
On the other hand, since Bugsplorer is trained with a line-level defect prediction objective, it can focus more on individual lines while making a prediction.
Thus, Bugsplorer can better pinpoint the defective lines and rank them higher in the list of suspicious lines.

\section{Methodology}
\label{sec: def-method}

\begin{figure*}
    \centering
    \includegraphics[width=.9\linewidth]{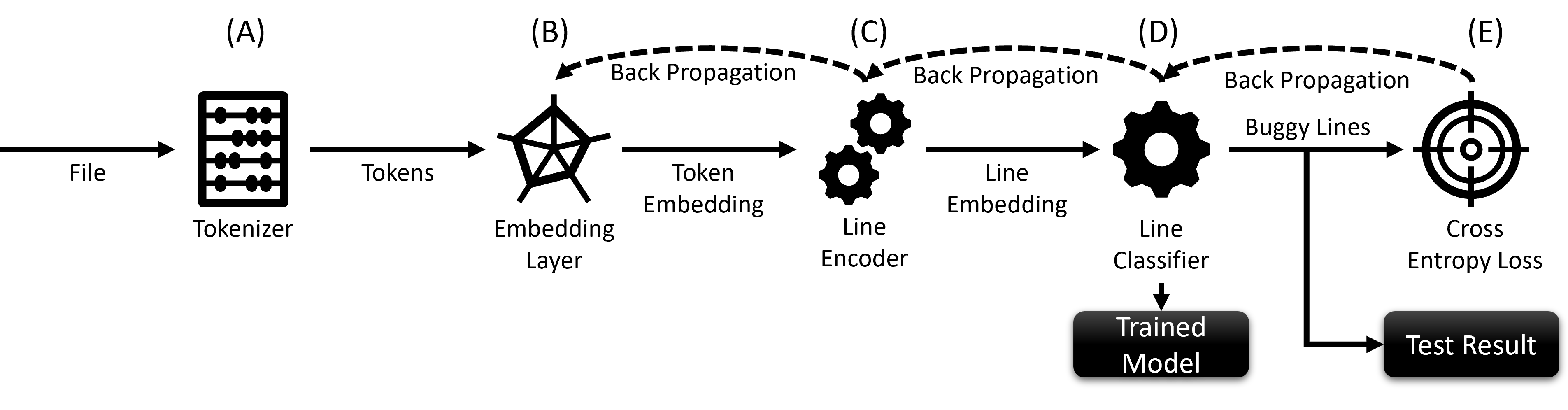}
    \caption{Schematic diagram of Bugsplorer}
    \label{fig: bugsplorer-schema}
    \vspace{-2em}
\end{figure*}

Fig.~\ref{fig: bugsplorer-schema} shows the schematic diagram of our proposed technique -- \emph{Bugsplorer} -- for predicting defects at the line level.
We discuss different steps of our technique in detail as follows.

\subsection{Pre-processing and Tokenization}
\label{subsec: def-preprocess}
Unlike many deep-learning (DL) models trained on code that treat source code documents as a stream of tokens~\cite{wattanakriengkrai2020predicting, pornprasit2021jitline, feng2020codebert, wang2021codet5}, we capture the hierarchical structure of source code documents (i.e., tokens forming lines and lines forming files).
We split each source code document into lines and represented them as a list of strings, where each string denotes a source code line (Fig.~\ref{fig: bugsplorer-schema}, Step A).
Then, we use a Byte-Pair Encoder (BPE) tokenizer~\cite{sennrich2016neural} to convert each line into distinct tokens.
BPE is a tokenizer that attempts to map a token to the largest possible sub-word in the vocabulary and falls back to smaller sub-words and even to a single letter in the case of rare words.

\looseness=-1
After the encoding, we represent each source code document as an integer matrix of shape $(L,\: T)$, where $L$ is the maximum number of lines in a file and $T$ is the maximum number of tokens in a line.
If a file has more lines than $L$, then we split the file into multiple entries with $N_O$ lines of overlap.
For example, if a file has $2L - N_O$ lines, we make one split from line 1 to $L$ and another from line $L - N_O + 1$ to line $2L - N_O$.
On the contrary, if a file has fewer lines than $L$, we fill it with lines containing only padding tokens.

\begin{figure}
    \centering
    \includegraphics[width=\linewidth]{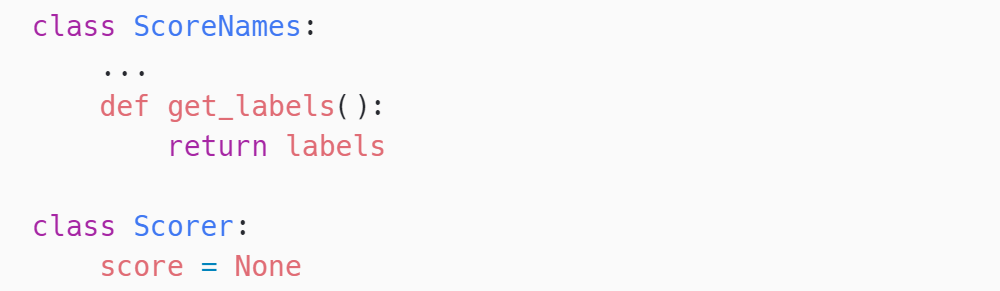}
    \caption{An example of the structural distance between two neighbouring tokens -- labels and Scorer}
    \label{fig: structural-distance}
    \vspace{-1em}
\end{figure}

\subsection{Token Embedding Generation}
\label{sec: def-embedding}
\looseness=-1
Bugsplorer uses both word embedding and positional embedding to represent the source code tokens (Fig.~\ref{fig: bugsplorer-schema}, Step B).
It starts with a word embedding layer that takes each source code document as an input and outputs a 3-dimensional matrix $(L,\: T,\: d_{model})$, where $d_{model}$ is the size of a vector representing the semantic information of a token.
Then, we pass this matrix to the positional embedding layer, which adds the positional information to the model.
The positional embedding informs the model which token comes after which.
In the original transformer model, the positional embedding was statically defined as a sinusoidal wave~\cite{vaswani2017attention}.
However, such a definition does not always reflect the structural distance between two tokens.
For instance, let us consider the code example in Fig.~\ref{fig: structural-distance}.
Here, the code tokens -- \texttt{labels} and \texttt{Scorer} -- belong to different class definitions.
Therefore, even though they are only two tokens apart, their structural distance is much larger.
Thus, to adapt to the structural aspects of source code documents in Bugsplorer, the positional embedding layer learns and optimizes the positional embedding of each token during the training phase.
Finally, similar to state-of-the-art transformer architectures (e.g., BERT~\cite{devlin2018bert}, RoBERTa~\cite{liu2019roberta}), we sum both word embedding and positional embedding and pass the new matrix of shape $(L, T, d_{model})$ to the line encoder.

\begin{figure}[t]
\centering
    \hfill%
    \begin{subfigure}[b]{.4\linewidth}
        \centering
        \includegraphics[width=\linewidth]{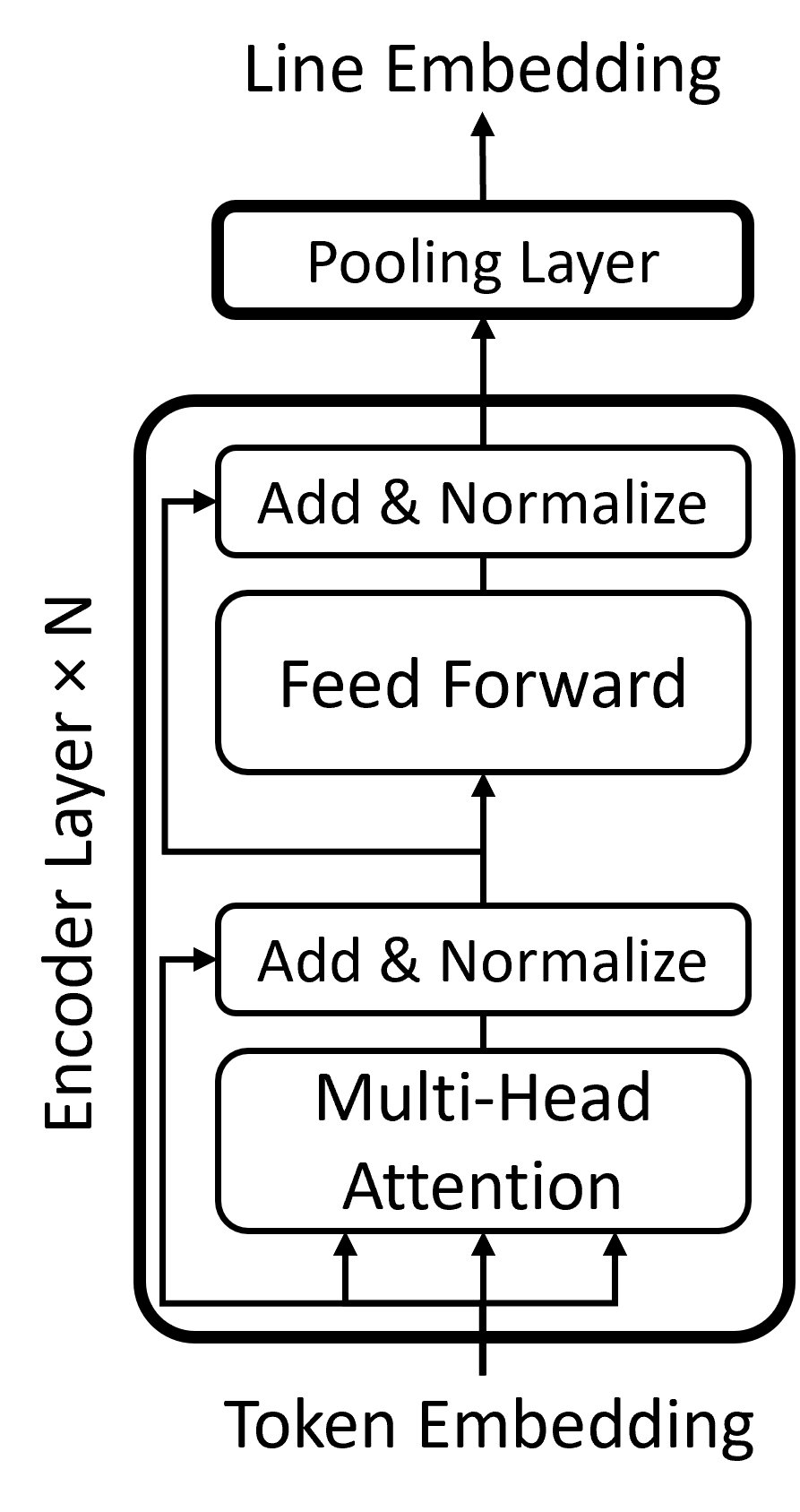}
        \caption{Line Encoder}
        \label{fig: line-encoder}
    \end{subfigure}
    \hfill%
    \begin{subfigure}[b]{.4\linewidth}
        \centering
        \includegraphics[width=.717\linewidth]{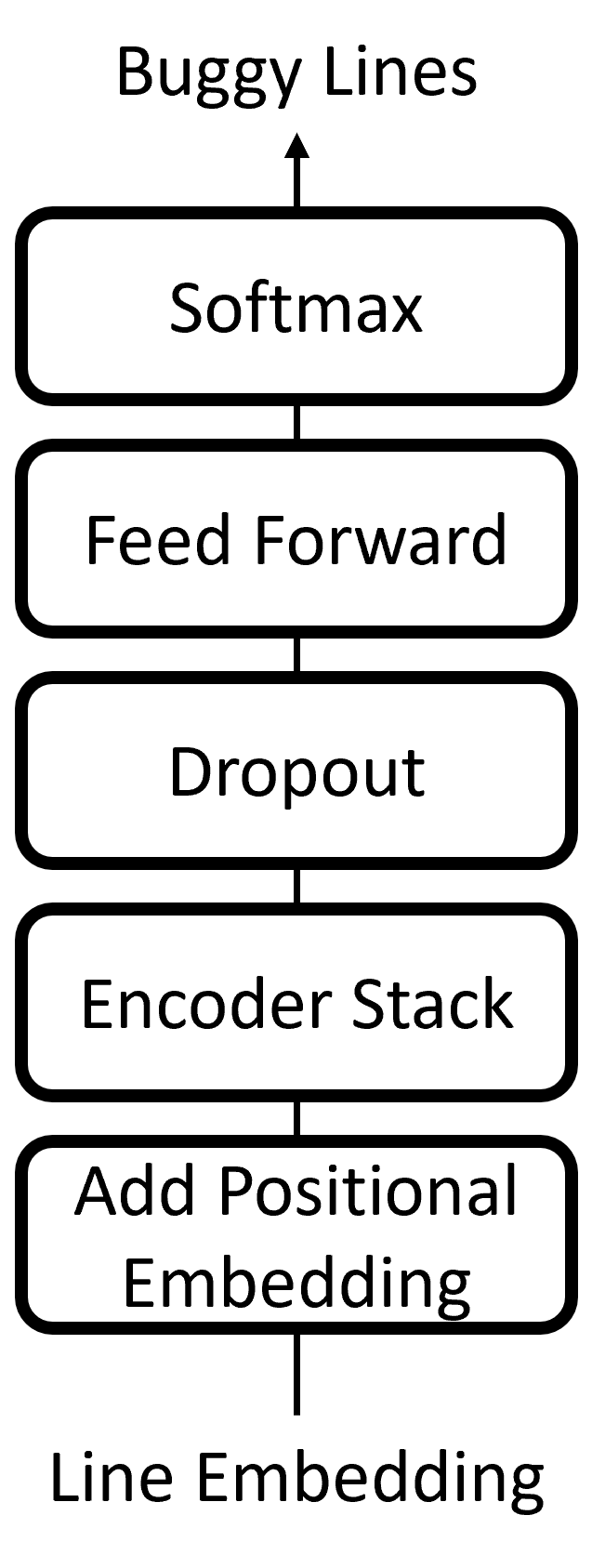}
        \caption{Line Classifier}
        \label{fig: file-encoder}
    \end{subfigure}
    \hfill%
    \begin{subfigure}[b]{.01\linewidth}
    \hspace{1mm}
    \end{subfigure}
    \caption{Internal architecture of the line encoder and line classifier}
    \vspace{-2em}
\end{figure}

\subsection{Line Embedding Generation}
\label{sec: line-encoder}

In this step, we pass the matrix representing a source code document (with semantic and positional information) to a transformer network.
We call it the \emph{line encoder} (Fig.~\ref{fig: bugsplorer-schema}, Step C).
For each line, the line encoder outputs a $d_{model}$-dimensional vector representing the semantics of the line.

\looseness=-1
Fig.~\ref{fig: line-encoder} shows a high-level overview of the line encoder.
The encoder stack has $N$ identical layers. 
Each layer has two parts: a multi-head self-attention network and a position-wise fully connected feed-forward neural network. 
Inside the attention layer, we query each line with all of its individual tokens to find the most informative tokens.
Then, during the back-propagation phase, each token learns to attend to all other tokens to determine their relative importance within the same line.
We aimed to learn an optimized representation of each source code token for the objective -- line-level defect prediction.
The encoder stack outputs a matrix of shape $(L,\: T,\: d_{model})$.
This means, at this stage, we still have a vector representation of size $d_{model}$ for every token in the file.
Interestingly, the vector representations at this stage are aware of other tokens in the same line and their relative importance in predicting the defective lines.
After that, we pass these token representations to a feed-forward network to capture line-level representation, commonly known as the pooling layer.
Unlike most CNN models that use a fixed pooling method (e.g., max pooling or average pooling), most transformer models (e.g., RoBERTa, T5) use a feed-forward network as the pooling layer.
This layer takes the vectors representing all tokens in a line as input and produces a single vector representing the source code line.
During the training, this layer learns to extract important information to detect defective code lines.
Thus, for each file, the pooling layer outputs a matrix of shape $(L,\: d_{model})$, where each row is a vector representing the semantics of a line.

\subsection{Line Classification}
\looseness=-1
The \emph{line classifier} (Fig.~\ref{fig: bugsplorer-schema}, Step D) accepts the vector representation of each line and determines their defect-proneness.
Fig.~\ref{fig: file-encoder} shows a high-level overview of our Line Classifier module.
It starts with a positional embedding layer that adds the positional information of \emph{each line} to their line embedding.
Similar to the positional embedding of tokens in the standalone Embedding Layer (i.e., Fig.~\ref{fig: bugsplorer-schema}, Step B), the positional embeddings of lines are also learned during the training phase.
The positional embedding layer is followed by the same encoder stack as that of the line encoder.
This encoder stack accepts the line embeddings as an input and outputs a new representation of the source code lines.
In particular, the encoder stack applies self-attention to the whole source document.
That is, each line attends to every other line to determine their relative importance within the same document.
Our goal was to find an optimized representation of each line by capturing not only their local but also global contexts.
This encoder stack has the same structures and hyper-parameters as the line encoder; thus, the details were skipped for brevity.
Then, the output of the encoder stack is passed to a feed-forward network via a dropout layer.
This feed-forward network outputs two values for each line indicating whether the line is defective or not.
Finally, we pass these values to a softmax layer, which performs a non-linear transformation to ensure the sum of two corresponding values is always 1.
Finally, we have a matrix of shape $(L,\: 2)$, indicating the probability of each line being defective and defect-free.

\subsection{Optimization}
\looseness=-1
After every training run, we identify the number of mistakes the model makes using a loss function.
Then, an optimizer algorithm identifies the nodes responsible for the mistakes and adjusts their weight accordingly.
We use cross-entropy loss~\cite{dwork1999mathematics} and AdamW optimizer~\cite{loshchilov2017decoupled} (Fig.~\ref{fig: bugsplorer-schema}, Step E).
The cross-entropy loss is defined as the number of bits needed to express the difference between two probability distributions (e.g., ground truth and prediction).
AdamW is an improvement over the more common Adam optimizer~\cite{kingma2014adam}.
The main difference between AdamW and Adam is how they implement regularization (i.e., preventing the model from overfitting).
AdamW enables a model to optimize some parameters while keeping the others unchanged.
Such optimization has been shown to lead a model to faster convergence and improved generalization performance~\cite{loshchilov2017decoupled}.
The amount of adjustment is dictated by a hyper-parameter named learning rate.
A large learning rate may prevent reaching the minimum loss, while a small one slows down the training.
Thus, we also use a linear scheduler to reduce the learning rate over time.
Existing baseline models like RoBERTa~\cite{liu2019roberta} and CodeT5~\cite{wang2021codet5} also used a linear scheduler to reduce their learning rate over time, which might justify our choice.

\section{Experiment}
\label{sec: def-exp}
We evaluate Bugsplorer with two large datasets constructed from 9 Java and 24 Python projects.
We examine its classification performance as well as its ability to rank the defective lines higher.
To place our work in the literature, we also compare our work with the existing state-of-the-art technique for line-level defect prediction~\cite{pornprasit2022deeplinedp}.
In our experiments, we thus answer four research questions as follows.

\begin{itemize}
    \item \textbf{RQ\textsubscript{1}}: How does Bugsplorer perform at line-level defect prediction in terms of classification performance and cost-effectiveness?
    
    \item \textbf{RQ\textsubscript{2}}: How do (a) the bidirectional representation of code elements (tokens and lines) and (b) the optimization of the model to line-level defect prediction affect Bugsplorer's performance?
    
    \item \textbf{RQ\textsubscript{3}}: How does the choice of transformer architecture affect the performance of Bugsplorer?
    
    \item \textbf{RQ\textsubscript{4}}: Can Bugsplorer outperform the existing state-of-the-art technique in terms of classification performance and cost-effectiveness?
\end{itemize}

\subsection{Experimental Datasets}

To evaluate Bugsplorer, we use two benchmark datasets -- Defectors~\cite{mahbub2023defectors} and LineDP~\cite{wattanakriengkrai2020predicting}.
Table~\ref{tab: datasets} provides the summary statistics of our benchmark datasets.

\looseness=-1
\textbf{Defectors} dataset contains source code documents and their defect locations from 24 popular Python systems across 18 domains and 24 organizations.
Unlike many existing datasets that use heuristics (e.g., issue number in commit messages) to identify bug-fixing changes, Defectors uses direct labelling of bug-fixing commits from the authors of the source code.
Then, the authors of the dataset identify corresponding defect-inducing changes using the popular SZZ algorithm~\cite{sliwerski2005szz}, followed by five levels of noise filtration recommended in the literature.
It contains a total of $\approx$~213K source code documents where $\approx$~93K are defective and $\approx$~120K are defect-free.
The Defectors dataset provides the dataset in two different splitting strategies -- \emph{random} and \emph{time-wise}.
It keeps 10K files for both validation and testing and the remaining $\approx$~193K files for training.

\textbf{LineDP} dataset contains 32 releases from 9 Java-based open-source software systems.
Each release contains 731 -- 8K files, 74K -- 567K lines of code, and 58K -- 621K code tokens.
All bug reports were retrieved from the JIRA Issue Tracking System (ITS) for each system.
Then, the authors of the dataset collect the bug-fixing changes associated with each bug-reporting issue.
They also used the SZZ algorithm~\cite{sliwerski2005szz} to identify defect-inducing changes from the bug-fixing changes.
‌\textcite{leclair2019recommendations} suggest that the training set should contain instances older than the testing set for an unbiased evaluation.
Thus, we keep the last release for each software system (total of 9) for testing, the second last release for each system (total of 9) for validation, and the remaining early releases (total of 14) for training.
This provides $\approx$~19K files for training, $\approx$~10K for validation, and $\approx$~24K for testing.

\begin{table}
    \centering
    \caption{Summary of the benchmark datasets}
    \label{tab: datasets}
    \begin{tabular}{|l|l|l|}
\hline
\textbf{Dataset}                        & \textbf{Defectors}    & \textbf{LineDP} \\ \hline \hline
\textbf{\# Files}                               & 213,419       & 73,395        \\ \hline
\textbf{\# Defective Files}                     & 93,668 (44\%) & 4,092 (6\%)   \\ \hline
\textbf{\# Defect-Free Files}                   & 119,751 (56\%)& 69,303 (94\%) \\ \hline
\textbf{Defective Lines in Defective Files}     & 4\%           & 0.34\%        \\ \hline
\end{tabular}
\vspace{-1em}
\end{table}

\subsection{Evaluation Metrics}
We evaluate Bugsplorer both as a classification and a retrieval technique using five appropriate performance metrics from the literature~\cite{pornprasit2022deeplinedp, wattanakriengkrai2020predicting, hoang2019deepjit} as follows.

\subsubsection{Balanced Accuracy}
Traditional accuracy measure is often biased toward the majority class~\cite{pornprasit2022deeplinedp}.
Balanced accuracy mitigates the problem by putting equal weight on the true positive result and the true negative result~\cite{urbanowicz2015exstracs}.

\subsubsection{Area Under the Receiver Operating Characteristic}
\looseness=-1
AuROC measures how well a model can discriminate between two different classes.
The receiver operating characteristic curve is the ratio between the true positive result and the false positive result~\cite{ferri2009experimental}.
AuROC is the area under this curve. 

\subsubsection{Recall@Top20\%LOC}
\looseness=-1
It measures the ratio between the number of defective lines in the top 20\% suspicious lines (i.e., with high defect-proneness) and the total number of defective lines~\cite{pornprasit2022deeplinedp}.
A value of 1.00 for Recall@Top20\%LOC means that all defective lines can be found within the top 20\% suspicious lines marked by a technique.
Assuming all defective lines are distributed naturally, a random guessing model will achieve a score of 0.20 for this metric.
A metric value higher than 0.20 indicates that the defective lines are concentrated at the top-ranked positions and one can find more defective lines with less effort.

\subsubsection{Effort@Top20\%Recall}
It measures the ratio between the number of suspicious lines that we have to investigate to find 20\% of the defective lines and the total number of ranked lines~\cite{pornprasit2022deeplinedp}.
A value of 1.00 for Effort@Top20\%Recall means that to find all defective lines, all the lines from the ranked list need to be investigated.
Assuming all defective lines are distributed naturally, a random guessing model will achieve a 0.20 score for this metric.
A \emph{lower} metric value indicates that one needs to put less effort into finding the defective lines.

\subsubsection{Initial False Alarm}
The initial false alarm (IFA) metric is the ratio between the number of misclassifications before the first true-positive and the total number of instances.
A lower value of IFA indicates that one needs to put less effort into finding the defective lines.

‌\subsection{Experiment Design and Hyper-Parameters}
\paragraph{Tokenizer}
We use a Byte-Pair Encoder (BPE) tokenizer that is pre-trained on GitHub CodeSearchNet~\cite{husain2019codesearchnet} dataset.
The dataset contains $\approx$~6M code snippets accompanied by documentation.
Since the tokenizer is trained on code corpus (as opposed to natural language corpus), it encodes source code with 33-50\% shorter length, compared to that of GPT2~\cite{radford2019language} or RoBERTa~\cite{liu2019roberta} tokenizer.

\paragraph{Encoder}
\looseness=-1
For each encoder stack in \emph{Line Encoder} and \emph{Line Classifier}, we use RoBERTa~\cite{liu2019roberta} transformer architecture.
Through experiments, we find that RoBERTa performs better for our research problem than other similar models (see Section~\ref{sssec: def-rq3}).
First, we initialize the learnable parameters from the encoder stack of the \emph{Line Encoder} using CodeBERTa pre-trained model from huggingface\footnote{https://huggingface.co/huggingface/CodeBERTa-small-v1}.
Similar to our tokenizer, this model too is pre-trained with the CodeSearchNet dataset.
Second, we initialize the learnable parameters of our second network, the \emph{Line Classifier}, using random values from a normal distribution with $\mu$ = 0 and $\sigma$ = 0.02 (same as RoBERTa).

\paragraph{Hyper-Parameters} 
\looseness=-1
We set the maximum number of lines in a file to 512 (i.e., $L = 512$) as the threshold.
While splitting large files into multiple parts, we use 64 lines of overlap (i.e., $N_O = 64$).
We make our train-validation-test datasets at the file level; thus, multiple splits of the same file reside in the same dataset.
This way, we ensure that the training dataset does not overlap with the validation or test dataset.


\paragraph{Hardware}
Our experiments are run on two NVidia A100 GPUs with 40GB of memory each. 
We use batches of $16$ files in each step (i.e., $16 \text{ files} \times 512 \text{ lines} \times 16 \text{ tokens} = 131,072$ tokens).
The average model training time is two days for the Defectors dataset and one day for the LineDP dataset.
The average evaluation time is $\approx$~12 minutes for the Defectors dataset (i.e., $\approx$~72 milliseconds per file) and $\approx$~25 minutes for the LineDP dataset (i.e., $\approx$~62 milliseconds per file).


\subsection{Evaluating Bugsplorer}

\subsubsection{\textbf{Answering RQ\textsubscript{1}} -- Performance of Bugsplorer}
\begin{figure}[t]
    \centering
    \includegraphics[width=\linewidth]{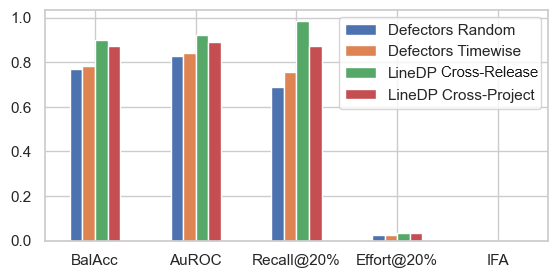}
    \vspace{-1.5em}
    \caption{Automated metric scores of Bugsplorer}%
    \label{fig: rq1}%
    \vspace{-2em}
\end{figure}

In this experiment, we evaluate Bugsplorer using five metrics in two different aspects -- classification and cost-effectiveness.
Fig.~\ref{fig: rq1} and Table~\ref{tab: rq1} show the performance of Bugsplorer.

First, we evaluate the performance of Bugsplorer using the random split variant of the Defectors dataset~\cite{mahbub2023defectors}.
For this dataset, Bugsplorer achieves a balanced accuracy of 0.77 and an AuROC of 0.83.
Such scores indicate a good capability of our technique in distinguishing true positive instances (i.e., defective lines) from true negative instances (i.e., defect-free lines).
In the case of cost-effectiveness metrics, Bugsplorer achieves a recall@20\%LOC score of 0.69.
Such a score means that with the help of our technique, an SQA engineer can find 69\% of all defective lines by only investigating 20\% lines from the ranked list.
Similarly, Bugsplorer archives 0.025 for effort@20\%recall, which means that to find 20\% of all defective lines, an SQA engineer needs to investigate only 2.5\% lines from the ranked list.
Finally, an initial false alarm (IFA) score of $\approx 0.00$ indicates a minimal effort to find the first true-positive instance (i.e., defective line).

\begin{table}[t]
    \centering
    \caption{Performance metric scores of Bugsplorer}%
    \label{tab: rq1}%
    \begin{tabular}{|l|r|r|r|r|}
    \hline
    \textbf{Metric} 
        & \parbox[t][2.5em]{12mm}{\textbf{Defectors Random}} 
        & \parbox[t][2.5em]{13mm}{\textbf{Defectors Timewise}} 
        & \parbox[t][3.5em]{11mm}{\textbf{LineDP Cross-Release}}
        & \parbox[t][3.5em]{11mm}{\textbf{LineDP Cross-Project}} \\
    \hline \hline
    BalAcc $\uparrow$           & 0.769 & 0.784 & 0.901 & 0.872 \\ \hline
    AuROC $\uparrow$ & 0.829 & 0.841 & 0.920 & 0.892 \\ \hline
    Recall@20\% $\uparrow$      & 0.690 & 0.754 & 0.985 & 0.871 \\ \hline
    Effort@20\% $\downarrow$    & 0.025 & 0.027 & 0.037 & 0.036 \\ \hline
    IFA $\downarrow$ & 0.000 & 0.000 & 0.006 & 0.004 \\ \hline
    \end{tabular}
    \newline \newline
    \footnotesize{ 
    $^*$ Up arrow ($\uparrow$) indicates higher is better and down arrow ($\downarrow$) indicates lower is better.
    }
    \vspace{-2em}
\end{table}

\looseness=-1
Even though the above experiment with a random split of the Defectors dataset shows promising results, we further evaluate Bugsplorer with the timewise variants of both datasets.
In particular, we leverage the timewise variant of the Defectors dataset and the cross-release variant of the LineDP dataset for our experiment.
While both test and validation sets from the timewise variant of Defectors contain the recently changed files from all projects, in the cross-release variant of LineDP, they contain the latest release from each project (inherently making it a timewise split).
From Table~\ref{tab: rq1}, we see that Bugsplorer performs well in timewise settings as well, interestingly, even better in some cases.
Bugsplorer achieves balanced accuracy scores between 0.78 to 0.90, and AuROC scores between 0.84 to 0.92.
Such scores indicate that Bugsplorer achieves high classification performance in timewise settings.
Our technique is cost-effective in timewise settings as well.
It achieves a recall@20\%LOC of 0.99 for the LineDP cross-release dataset.
This means that one can find nearly all defective lines by checking only the top 20\% suspicious lines from Bugsplorer.
Bugsplorer scores between 0.027 and 0.037 for effort@20\%recall, which means that an SQA engineer needs to investigate only 2.7\%--3.7\% suspicious lines to find 20\% of the defective lines.
Finally, an initial false alarm (IFA) score of $\approx$ 0.00--0.01 indicates a minimal overhead to find the first defective line on the ranked list.
Interestingly, even though ML models tend to perform better with randomly split data~\cite{mahbub2022explaining,tao2021evaluation}, the performance of Bugsplorer in both random and timewise splits of the Defectors dataset is comparable.
Such a phenomenon indicates the robustness of our technique at line-level defect prediction with unseen data.

\looseness=-1
In practical scenarios, when applying Bugsplorer to a new project, obtaining project-specific data and retraining our model may not be possible. 
To simulate such scenarios, we also evaluate the performance of Bugsplorer in a cross-project setting with LineDP. 
This means that the training, validation, and testing datasets contain commits from entirely separate projects, ensuring mutual exclusivity.
To avoid bias towards any particular project, we created nine variants of the LineDP dataset for cross-project setting.
In each variant, we take one project for validation, one for testing, and the remaining seven projects for training.
Such a setting ensures that Bugsplorer is tested with each project.
Then, we report the average score from each variant.
From Table~\ref{tab: rq1}, we see that Bugsplorer shows a mixed trend in performance with cross-project setting.
For the balanced accuracy, AuROC, and recall@20\%LOC metrics, the performance drops by 3\%, 3\%, and 12\%, respectively.
However, the metric scores achieved by Bugsplorer in cross-project settings are still promising (e.g., 0.89 AuROC).
Interestingly, for effort@20\%recall and initial false alarm (IFA) metrics, our technique performs 3\% and 33\% better in cross-project settings, respectively.
Thus, overall, Bugsplorer can significantly reduce the costs of finding defects even in cross-project setting.

\begin{tcolorbox}[left=2pt,right=2pt,top=2pt,bottom=2pt]
\textbf{Summary of RQ\textsubscript{1}:} Bugsplorer shows promising results at line-level defect prediction with a balanced accuracy of up to 0.90 and an AuROC of up to 0.92. It can also rank the first 20\% of the defective lines within the top 2-3\% of its suspicious lines, which is promising. 
\end{tcolorbox}

\begin{table*}
\centering
\caption{Effectiveness of Bi-directional Representation of Code Elements and Line-Level Optimization}
\label{tab: rq2}
\begin{tabular}{|l|l|p{12.75mm}|l||l|p{12.75mm}|l||l|p{12.75mm}|l|}
\hline
\textbf{Dataset} & \multicolumn{3}{c||}{\textbf{Defectors Random}} & \multicolumn{3}{c||}{\textbf{Defectors Timewise}}   & \multicolumn{3}{c|}{\textbf{LineDP Cross-Release}}    \\ \hline
\textbf{Technique}               & Bugsplorer    & Bugsplorer$_F$ & DeepLineDP   & Bugsplorer   & Bugsplorer$_F$   & DeepLineDP    & Bugsplorer    & Bugsplorer$_F$   & DeepLineDP  \\ \hline \hline
\textbf{BA $\uparrow$}           & 0.769         & 0.603         & 0.610          & 0.784        & 0.628             & 0.561         & 0.901         & 0.605             & 0.538       \\ \hline
\textbf{AuROC $\uparrow$}        & 0.829         & 0.610         & 0.633          & 0.841        & 0.630             & 0.518         & 0.920         & 0.556             & 0.510       \\ \hline
\textbf{Recall@20\% $\uparrow$}  & 0.690         & 0.320         & 0.324          & 0.754        & 0.380             & 0.281         & 0.985         & 0.251             & 0.224       \\ \hline
\textbf{Effort@20\% $\downarrow$}& 0.025         & 0.111         & 0.089          & 0.027        & 0.085             & 0.105         & 0.037         & 0.167             & 0.191       \\ \hline
\textbf{IFA $\downarrow$}        & 0.000         & 0.000         & 0.002          & 0.000        & 0.000             & 0.000         & 0.006         & 0.006             & 0.007       \\ \hline       
\end{tabular}
\vspace{-2em}
\end{table*}

\subsubsection{\textbf{Answering RQ\textsubscript{2}} -- Effectiveness of Bi-directional Representation of Code Elements and Line-Level Optimization}
\label{sssec: rq2}
\looseness=-1
In this experiment, we analyze the effectiveness of (a) using bidirectional representations of code elements (e.g., tokens and lines) instead of concatenating two unidirectional representations and (b) line-level optimization during model training.
First, we introduce a new variant of Bugsplorer -- Bugsplorer$_{F}$, which is trained with the objective of file-level defect prediction.
Then, we compare (a) Bugsplorer$_{F}$ and DeepLineDP~\cite{pornprasit2022deeplinedp} to determine the effectiveness of bidirectional representation and (b) Bugsplorer$_{F}$ and Bugsplorer to determine the effectiveness of line-level optimization during model training.
Table~\ref{tab: rq2} shows the performances of Bugsplorer, Bugsplorer$_{F}$, and DeepLineDP.
Fig.~\ref{fig: rq2} illustrates their performances using boxplots.
Since increments are desirable for some metrics and decrements are desirable for others, we use the terms -- better performance or worse performance -- to explain them.
We also mark them using up-arrow and down-arrow in Table~\ref{tab: rq2} respectively.

\looseness=-1
Bugsplorer$_{F}$ uses a transformer network to encode source code elements (e.g., tokens or lines), whereas DeepLineDP~\cite{pornprasit2022deeplinedp} uses a Recurrent Neural Network (RNN).
The use of a transformer network lets Bugsplorer$_{F}$ focus on surrounding tokens from both sides of a token simultaneously, leading to bidirectional representations of the code elements.
On the contrary, using RNN, DeepLineDP generates two unidirectional representations of each line (i.e., one is from left to right, and the other is from right to left) and then concatenates them to generate a representation of the lines.
Thus, a comparison between Bugsplorer$_{F}$ and DeepLineDP can reveal the effectiveness of our bidirectional representations for code elements.
From Table~\ref{tab: rq2}, we see that in most cases, Bugsplorer$_{F}$ shows better performance than that of DeepLineDP.
For the timewise split of Defectors, Bugsplorer$_{F}$ shows 12--35\% better scores in balanced accuracy, AuROC, recall@20\%LOC, and effort@20\%recall metrics.
Similarly, for the LineDP dataset, Bugsplorer$_{F}$ shows 9--15\% better performance in all metrics.
Finally, we see a mixed trend for the random split of Defectors.
DeepLineDP shows a 1-3\% better performance for balanced accuracy, AuROC, and recall@20\%LOC metrics, which are marginally better.
For the effort@20\%recall metric, DeepLineDP archives 24\% better performance (actual metric score reduced by only 0.022).
Nonetheless, for the initial false alarm metric (lower is better), the score of DeepLineDP increased from $\approx$~0.0 to 0.002.
This means that to find the first defective lines with DeepLineDP, one has to investigate 0.2\% lines of the ranked list, whereas the amount is $\approx$~0\% for Bugsplorer$_{F}$.
Given the evidence above, our choice of generating bidirectional representations for code elements (e.g., lines or tokens) using a transformer network might be justified.

While Bugsplorer$_{F}$ is trained with a file-level defect prediction objective, Bugsplorer is trained with a line-level defect prediction objective.
However, they share the same network architecture apart from their output layer.
Therefore, a comparison between them can reveal the effectiveness of their optimization level during model training.
Table~\ref{tab: rq2} shows that Bugsplorer outperforms Bugsplorer$_{F}$ in nearly all metric scores across all datasets.
For balanced accuracy, Bugsplorer shows 25--49\% better performance, while for AuROC, the improvement is 33--65\%.
Such improvements indicate that the line-level optimization during model training (i.e., Bugsplorer) leads to better classification performance with a strong capability of discriminating between defective and defect-free lines.
In cost-effectiveness metrics, we see even bigger improvements.
The line-level optimization in defect prediction achieves 98--292\% better scores in terms of recall@20\%LOC.
Similarly, the effort@20\%recall score is 68--78\% better.
Finally, the initial false alarm score is the same for both variants across all datasets. 
All these improvements in metric scores suggest that line-level optimization is a much better choice than file-level optimization during model training, which justifies our choice.

\begin{tcolorbox}[left=2pt,right=2pt,top=2pt,bottom=2pt]
\textbf{Summary of RQ\textsubscript{2}:} Both the bi-directional representation of code elements and the line-level defect prediction objective lead to better performance in our technique. Given all the evidence above, our choices regarding token representation and optimization level might be justified. 
\end{tcolorbox}

\subsubsection{\textbf{Answering RQ\textsubscript{3}} -- Impact of the Choice of Transformer Architecture on Bugsplorer}
\label{sssec: def-rq3}

In this experiment, we investigate how our choice of the transformer architecture in the \emph{encoder stack} affects the performance of Bugsplorer.
In particular, we experiment with three popular transformer architectures -- RoBERTa~\cite{liu2019roberta}, BERT~\cite{devlin2018bert}, and T5~\cite{raffel2020exploring}, because of their extensive use in the software engineering domain and state-of-the-art performances with relevant benchmarks like CodeSearchNet~\cite{husain2019codesearchnet} and CodeXGLUE~\cite{lu2021codexglue}.
To initialize the learnable parameters of Line Encoder (Section~\ref{sec: line-encoder}), we use CodeBERT~\cite{feng2020codebert} for BERT, CodeBERTa\footnote{https://huggingface.co/huggingface/CodeBERTa-small-v1} for RoBERTa, and CodeT5~\cite{wang2021codet5} for T5.
All of these models were pre-trained with the CodeSearchNet dataset.
We initialize the learnable parameters of the Line Classifiers using normally distributed random values in all variants.
Note that even though a T5 model contains both an encoder and a decoder, we use only the encoder part in our work.
Table~\ref{tab: rq3} shows the performance of Bugsplorer with these three transformer architectures.
Since increments are desirable for some metrics and decrements are desirable for others, we use the terms -- better performance or worse performance -- to explain them.
We also mark them using up-arrow and down-arrow in Table~\ref{tab: rq3} respectively.

\begin{table*}
\centering
\caption{Performance of Bugsplorer with different transformer architectures}
\label{tab: rq3}
\begin{tabular}{|l|l|l|l||l|l|l||l|l|l|}
\hline
\textbf{Dataset} & \multicolumn{3}{c||}{\textbf{Defectors Random}} & \multicolumn{3}{c||}{\textbf{Defectors Timewise}}   & \multicolumn{3}{c|}{\textbf{LineDP Cross-Release}}    \\ \hline
\textbf{Architecture}               & RoBERTa   & BERT  & T5    & RoBERTa   & BERT  & T5    & RoBERTa   & BERT  & T5    \\ \hline \hline
\textbf{BA $\uparrow$}              & 0.769     & 0.769 & 0.709 & 0.784     & 0.778 & 0.710 & 0.901     & 0.849 & 0.909 \\ \hline
\textbf{AuROC $\uparrow$}           & 0.829     & 0.828 & 0.795 & 0.841     & 0.845 & 0.791 & 0.920     & 0.897 & 0.914 \\ \hline
\textbf{Recall@20\% $\uparrow$}     & 0.690     & 0.690 & 0.572 & 0.754     & 0.754 & 0.577 & 0.985     & 0.871 & 0.995 \\ \hline
\textbf{Effort@20\% $\downarrow$}   & 0.025     & 0.025 & 0.029 & 0.027     & 0.027 & 0.036 & 0.037     & 0.037 & 0.034 \\ \hline
\textbf{IFA $\downarrow$}           & 0.000     & 0.000 & 0.000 & 0.000     & 0.000 & 0.000 & 0.006     & 0.006 & 0.006 \\ \hline       
\end{tabular}
\vspace{-2em}
\end{table*}

When comparing RoBERTa with BERT, there is no clear winner.
In most cases, they achieve nearly the same performance.
Even when their scores differ, the difference is only marginal in most cases.
In particular, for the random split of Defectors, both of them achieve the same scores for balanced accuracy, recall@20\%LOC, effort@20\%recall, and initial false alarm metrics.
Only for the AuROC metric, RoBERTa shows 0.2\% worse performance, which is marginal.
For the timewise split of Defectors, the performance of RoBERTa varies from 0.5\% worse to 0.8\% better in balanced accuracy, AuROC, recall@20\%LOC, and effort@20\%recall metrics.
For the initial false alarm, the score remains the same.
For the LineDP cross-release dataset, RoBERTa shows 2--12\% better performance in balanced accuracy, AuROC, and recall@20\%LOC metrics.
Both architectures achieve the same performance for effort@20\%recall and initial false alarm.
Considering the trend in these metric scores, we see that the performance of RoBERTa is marginally better than that of BERT.
Since the RoBERTa model is a successor of the BERT model while sharing almost similar internal architectures, such a trend in their performances might be expected. 

\looseness=-1
When comparing RoBERTa with T5, RoBERTa consistently performs better than T5 for both variants of the Defectors dataset but shows dissimilar patterns for the LineDP cross-release dataset.
For the random split of Defectors, RoBERTa shows 4--17\% better performance in balanced accuracy, AuROC, recall@20\%LOC, and effort@20\%recall metrics.
For the timewise split of Defectors, RoBERTa consistently shows better performance (6--34\%) for balanced accuracy, AuROC, recall@20\%LOC, and effort@20\%recall metrics.
However, for the LineDP cross-release dataset, we see some mixed trends.
RoBERTa shows 1\% worse performance in balanced accuracy and recall@20\%LOC metrics while achieving 1\% better performance for the AuROC metric.
Nonetheless, for the effort@20\%recall metric, RoBERTa shows 9\% worse performance.
Finally, for the initial false alarm, both of the architectures perform the same across all datasets.
Thus, T5 and RoBERTa show mixed performance trends in the LineDP dataset, whereas T5 consistently performs worse in the Defectors dataset.
Since T5 is designed for both encoding and decoding, whereas RoBERTa is specialized for encoding, such performance differences among them might be explainable.

Based on the evidence shown above, we find that even though the use of transformer architecture is crucial (see Section~\ref{sssec: rq2}), its type has minimal impact on the performance of Bugsplorer. 
RoBERTa performs marginally better than the others.
Therefore, we use RoBERTa as the \emph{default choice} of Bugsplorer.

\begin{tcolorbox}[left=2pt,right=2pt,top=2pt,bottom=2pt]
\textbf{Summary of RQ\textsubscript{3}:} RoBERTa performs better than T5 across all datasets for nearly all metric scores. It also marginally performs better than BERT. 
Therefore, our choice of using RoBERTa as the default choice of Bugsplorer might be justified.
\end{tcolorbox}

\subsubsection{\textbf{Answering RQ\textsubscript{4}} -- Comparison with the Existing Baseline Technique}
\looseness=-1
In this research question, we compare Bugsplorer with the state-of-the-art technique for line-level defect prediction -- DeepLineDP~\cite{pornprasit2022deeplinedp}.
Since DeepLineDP outperforms all previous techniques, it can be considered the state-of-the-art technique for line-level defect prediction.
We use the replication package from original authors and evaluate DeepLineDP against both of our benchmark datasets for comparison.
We investigate whether Bugsplorer can outperform it in terms of classification performance and cost-effectiveness.

\begin{figure}
    \centering
    \includegraphics[width=.9\linewidth]{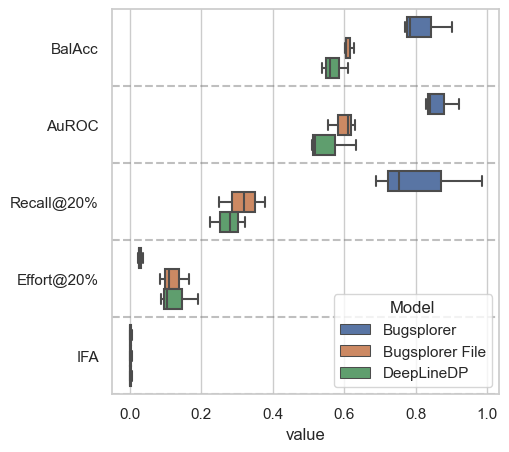}
    \vspace{-.5em}
    \caption{Effectiveness of Bidirectional Representation of Code Elements and Line-Level Optimization}
    \vspace{-2em}
    \label{fig: rq2}
\end{figure}

\looseness=-1
Table~\ref{tab: rq2} and Fig.~\ref{fig: rq2} compare between Bugsplorer and DeepLineDP across all datasets and all metrics. We see that Bugsplorer outperforms DeepLineDP in all aspects.
In the case of classification, Bugsplorer achieves 26-68\% better performance for balanced accuracy and 31--80\% better performance for AuROC.
We also see a similar trend in cost-effectiveness.
For recall@20\%LOC and effort@20\%recall metrics, Bugsplorer achieves 113--340\% and 72--81\% improved performance, respectively.
Such improvements indicate that our technique can significantly reduce the effort needed to find defective lines in a codebase.
Finally, for the initial false alarm metric, our technique shows 0--97\% better performance.
Thus, Bugsplorer outperforms DeepLineDP both in terms of classification capability and cost-effectiveness.

\looseness=-1
Similar to Bugsplorer, DeepLineDP uses a hierarchical structure of neural networks.
It uses two RNNs (inherently GRUs) to build the model, whereas Bugsplorer uses two transformer networks based on the RoBERTa architecture~\cite{liu2019roberta}.
Due to a sequential architecture like RNN, DeepLineDP can represent a line only unidirectionally, either from left to right or right to left.
Then it concatenates these two representations to make a bidirectional representation.
On the contrary, Bugsplorer can directly make a bidirectional representation of a line via the Line Encoder (Section~\ref{sec: line-encoder}).
Furthermore, during the training phase, Bugsplorer is optimized for line-level defect prediction, whereas DeepLineDP is optimized for file-level defect prediction.
Both of these \emph{novel contributions} (i.e., bidirectional representation and line-level optimization) are proven to be beneficial in RQ\textsubscript{2}.
Thus, Bugsploer's better performance than that of DeepLineDP is explainable.

\begin{tcolorbox}[left=2pt,right=2pt,top=2pt,bottom=2pt]
\textbf{Summary of RQ\textsubscript{4}:} Bugsplorer outperforms the state-of-the-art technique for line-level defect prediction. Bugsplorer is 26-68\% more accurate in predicting the defective lines from source code. It can also reduce the effort in finding defective lines by 72-81\%. 
\end{tcolorbox}

\section{Threats To Validity}
\label{sec: def-threats}

\textit{Threats to internal validity} relate to experimental errors and biases~\cite{mahbub2022explaining, shuvo2023revcom}.
Re-implementation of the existing techniques could pose a such threat.
However, while implementing the DeepLineDP technique~\cite{pornprasit2022deeplinedp}, we use the replication package provided by the authors.
Possible errors in the implementation of our technique could also pose a threat.
To avoid such errors, we carefully developed the technique with several rounds of revision followed by rigorous testing.
Therefore, the threats to the internal validity posed by Bugsplorer might be minimal.

\looseness=-1
\textit{Threats to construct validity} are factors that may affect how well a test or measure assesses what it is supposed to measure~\cite{mondal2019can}.
We use five evaluation metrics to evaluate Bugsplorer in both classification and cost-effectiveness aspects.
Given the severe class imbalance in datasets (less than $1\%$ defective lines), we chose the metrics minimally affected by class imbalance.
Furthermore, these metrics were also widely used by similar prior works~\cite{wattanakriengkrai2020predicting, pornprasit2021jitline, pornprasit2022deeplinedp}.
Since Bugsplorer only takes a single file as input, its capability of finding defects that span multiple files (e.g., incorrect API use) might pose a threat.
However, Bugsplorer learns to predict defective lines based on previous mistakes. 
Thus, it could detect such defects if the training dataset contains similar instances. 
In other words, even though Bugsplorer accepts single-file input, it could identify defects related to external files.
Nonetheless, we acknowledge that our technique might be limited in this regard.

\looseness=-1
\textit{Threats to external validity} relate to the generalizability of our technique~\cite{mahbub2022explaining, shuvo2023revcom}.
We evaluate Bugsplorer using two benchmark datasets~\cite{mahbub2023defectors, wattanakriengkrai2020predicting} constructed from Python and Java software systems.
These datasets contain 33 software systems in total.
Furthermore, the software systems in the Python dataset -- Defectors -- are from 18 application domains and 24 organizations.
Thus, our evaluation using these large and diverse datasets could mitigate the threats to external validity.
‌
\section{Manual Analysis}

In this section, we perform a qualitative analysis to investigate the scenarios where Bugsplorer shines and the scenarios where it struggles.
In particular, we categorized the predictions from Bugsplorer as true positives, true negatives, false positives, and false negatives.
Then, we analyze 100 random samples from each category to find patterns within them.
We summarize our findings below.

\subsubsection*{False Positives}
\looseness=-1
The most common pattern in this category is the use of long comments that look like code.
In particular, more than half of our samples (52) have comments spanning three or more lines.
Example 1 in Table~\ref{tab: manual-analysis} shows such a case where an IPython code example is added as a comment that spans eight lines.
Embedding structural information to the source code~\cite{mahbub2022explaining, shippey2019automatically} might mitigate such issues.
Another common pattern is the use of valid but rare syntax.
Declaring a class within a class is a valid but rarely used Python syntax.
Therefore, Bugsplorer might predict it as a defective line.

\subsubsection*{False Negatives}
The most common pattern in this category is the code that depends on the environment.
It is hard to know whether such code is defective or not just by looking at the code (a.k.a. extrinsic bug~\cite{rodriguez2020watch}).
Some common examples of such a pattern are reading environment variables or reading a file (Example 5).
Nearly one-fifth of our samples in this category (21) follow this pattern.
Another interesting pattern is code comments labelled as defective.
Even though, in most cases, code comments do not cause any defect, the benchmark datasets labelled them as defective in some cases.
Nonetheless, Bugsplorer can identify code comments and mark them as defect-free even though some training data says otherwise.

\begin{table}
    \caption{Examples of classification by Bugsplorer. A left arrow (\texttt{<---}) indicates the predicted buggy line.}
    \label{tab: manual-analysis}
    \centering
    \begin{tabular}{|p{3mm}|p{77mm}|}
        \hline
        \textbf{Eg} & \textbf{Code} \\ \hline \hline
        1\linebreak \linebreak \linebreak FP & 
        \begin{lstlisting}[language=Python]
>>> from torch import Tensor
>>> class ExampleModule(DeviceDtypeModuleMixin):
...     def __init__(self, weight: Tensor): <---
...         super().__init__()
...         self.register_buffer('weight', weight)
>>> _ = torch.manual_seed(0)
>>> module = ExampleModule(torch.rand(3, 4))
>>> module.weight #doctest: +ELLIPSIS
\end{lstlisting} \\ \hline

    2 \linebreak \linebreak \linebreak \linebreak TP & \begin{lstlisting}[language=Python]
# Now we need to find the missing filenames for the subpath we want.
# Looking for this 'rev-list' command in the git --help? Hah.
cmd = f"git -C {tmp_dir} rev-list --objects --all --missing=print -- {subpath}" <---
ret = run_command(cmd, capture=True)
\end{lstlisting} \\ \hline

    3 \linebreak \linebreak \linebreak \linebreak \linebreak TP & \begin{lstlisting}[language=Java]
try {
  // delete done file
  boolean deleted = operations.deleteFile(doneFileName); <---
  log.trace("Done file: {} was deleted: {}", doneFileName, deleted); <---
    if (!deleted) { <---
    log.warn("Done file: " + doneFileName + " could not be deleted"); <---
  }
} catch (Exception e) {
  handleException(e);
}
\end{lstlisting} \\ \hline

    4 TP & \begin{lstlisting}[language=Python]
properties.setProperty("user","cloud");
properties.setProperty("password","scape"); <---
\end{lstlisting} \\ \hline

    5 \linebreak \linebreak \linebreak \linebreak FN & \begin{lstlisting}[language=Python]
elif is_path:
    if compat.PY2:
        # Python 2
        f = open(path_or_buf, mode) <---
    elif encoding:
        # Python 3 and encoding
        f = open(path_or_buf, mode, encoding=encoding)
    else:
        # Python 3 and no explicit encoding
\end{lstlisting} \\ \hline

    \end{tabular}

    \vspace{-2em}
\end{table}

\subsubsection*{True Positives}
\looseness=-1
An interesting finding is that Bugsplorer has not only the ability to find bugs in various programming languages (e.g., Python or Java), but it also knows common tools (e.g., git) as well.
For instance, Example 2 shows a git command that uses the \texttt{missing=print} option which is added in version 2.22.
The fixed version of that code\footnote{https://bit.ly/3LnpaXj} also fixes the issue by checking whether the installed git version is $\ge$ 2.22 or not.
Another interesting finding is that Bugsplorer is quite precise in identifying consecutive defective lines (Example 3).
Bugsplorer is good at identifying security vulnerabilities as well.
Example 4 shows a case where the password is hardcoded, whereas it should be read from some configuration file.

\subsubsection*{True Negatives}
Unfortunately, it is hard to find any pattern within this category containing all defect-free code.

\section{Related Work}
\label{sec: def-related}

\subsection{Defect Prediction at Various Granularity Levels}
Defect prediction has been a popular research topic for the last few decades. 
Earlier works predicted defects at different granularity levels of code such as module~\cite{gong2021revisiting, yu2019empirical}, file~\cite{chen2020software, kamei2010revisiting}, method~\cite{shippey2019automatically, hata2012bug}, and commit~\cite{mcintosh2018fix, pascarella2019fine, huang2019revisiting, pornprasit2021jitline, hoang2019deepjit, hoang2020cc2vec, kamei2012large}.
Finding the actual lines of code that contain defects still consumes significant time and effort from developers. 
Two recent studies~\cite{wan2018perceptions, pornprasit2022deeplinedp} independently show that practitioners could benefit from fine-grained defect prediction such as line-level defect prediction. 
It can help developers focus their SQA efforts on the vulnerable parts of the source code.

\subsection{Defect Prediction with Machine Learning}
Machine learning-based approaches for defect prediction primarily rely on different metric scores to identify defective entities (e.g., file or commit).
\textcite{kamei2012large} perform a large-scale study on change-level defect prediction using six open-source and five closed-source projects.
They proposed a total of 14 metric scores to predict defects at the file level with a logistic regression model.
\textcite{mcintosh2018fix} conduct a time-series analysis on JIT defect prediction using two rapidly evolving projects.
They extracted 17 code properties and showed that the importance of these code properties in predicting the defective commits change over time.
\textcite{jiang2013personalized} attempt to personalize defect prediction for different developers.
They used bag-of-words and characteristics vector (i.e., count of each node type in AST) to predict the defects at the file level.
Even though these works lay the ground for further defect prediction research, they are often limited by their coarse granularity and ordinary performance.

\subsection{Defect Prediction with Deep Learning}
\looseness=-1
Previous deep learning-based defect prediction models used various architectures to extract semantic and syntactic features from source code.
\textcite{wang2016automatically} proposed a Deep Belief Network (DBF) architecture that represents a source code document using semantic features derived from the AST.
Li et al.~\cite{li2017software,li2019improving} proposed a CNN architecture that learns the semantic and structural features of source code documents from the token sequences and the AST, PDG and DFG. 
\textcite{dam2018automatic} and \textcite{zou2019mu} individually proposed a Long Short-Term Memory (LSTM) architecture that can learn the semantic and syntactic features of source code documents from the token sequences and the CFG. 
However, these models only predict defects at the file level, which is too coarse-grained. In contrast, our deep learning-based approach predicts defects at the line level and thus can identify defective lines of source code.

\subsection{Line-Level Defect Prediction}
\looseness=-1
Prior studies attempt to predict defects at the line level using various approaches, including static analysis and machine learning.
Static analysis tools produce too many false positive results~\cite{kamei2012large} as well as false negative results~\cite{thung2015extent}.
In the last few years, line-level defect prediction with an explainable model has been popular.
\textcite{wattanakriengkrai2020predicting} train a model to predict defects at the file level.
Then, they use a model explainer tool -- LIME~\cite{ribeiro2016should} -- to get importance values for each input (i.e., tokens).
Those importance values, in turn, are used to find defective lines within a file containing highly important tokens.
Later, \textcite{pornprasit2021jitline} adapted this technique to identify defective lines from commit diffs.
Recently, they proposed DeepLineDP\cite{pornprasit2022deeplinedp} that trains a GRU model~\cite{cho2014learning} with attention mechanism~\cite{bahdanau2014neural} to predict defects at the file level.
Then, they rank the lines with highly attended tokens as the candidate defective lines.

All the above approaches share two common limitations. 
First, their models learn with a file-level defect prediction objective.
As a result, when these values are later used to predict defects at the line level, their performance could be sub-optimal.
On the contrary, Bugsplorer directly learns to predict defects at the line level and thus can focus on a finer-grained context of each line.
Second, these techniques either use no contextual information (e.g., LineDP~\cite{wattanakriengkrai2020predicting}) or unidirectional context (e.g., DeepLineDP~\cite{pornprasit2022deeplinedp}), whereas Bugsplorer learns bidirectional representations of the code elements.
Both of these \emph{novel} improvements are shown to be more accurate and cost-effective (see Sec~\ref{sssec: rq2}).

\section{Conclusion and Future Works}
\label{sec: conclusion}
\looseness=-1
Software bugs not only claim precious development time but also cost billions every year.
In this study, we propose a novel transformer-based technique -- Bugsplorer -- to predict defects at the line level. 
Our evaluation with five performance metrics shows that our technique has a promising capability of predicting defective lines with 26-72\% higher accuracy than the state-of-the-art. 
It can rank the first 20\% defective lines within the top 1-3\% vulnerable lines.
Thus, Bugsplorer has the potential to significantly reduce SQA costs by ranking defective lines higher.

In future, we will investigate to make Bugsplorer more robust against rarely used syntax.
Furthermore, we will explore whether embedding structural information (e.g., AST, PDG) with the source code can improve defect prediction.

\balance

\scriptsize
\printbibliography

\end{document}